\documentclass[nofootinbib,preprint,tightenlines,superscriptaddress]{revtex4}

\usepackage[utf8]{inputenc}
\usepackage{amssymb,amsthm,amsmath,amstext,amsbsy}
\usepackage{url}
\usepackage{nicefrac}
\usepackage{graphicx}
\usepackage{hyperref}

\newcommand{\ie}{\textit{i.e.}}
\newcommand{\eg}{\textit{e.g.}}
\newcommand{\cf}{\textit{cf.}}

\newcommand{\etal}{\textit{et al.}}

\newcommand{\mathspace}{\ \ }
\newcommand{\mathtext}[1]{\mathspace\text{#1}\mathspace}

\newcommand{\keV}{\ensuremath{\mathrm{keV}}}
\newcommand{\MeV}{\ensuremath{\mathrm{MeV}}}

\newcommand{\fm}{\ensuremath{\mathrm{fm}}}

\newcommand{\vk}{\mathbf{k}}

\newcommand{\dd}{\mathrm{d}}

\newcommand{\ii}{\mathrm{i}}

\newcommand{\vD}{\boldsymbol{D}}
\newcommand{\hc}{\mathrm{h.c.}}
\newcommand{\OO}{\mathcal{O}}

\newcommand{\eps}{\varepsilon}

\newcommand{\MN}{M_N}

\newcommand{\yd}{y_d}
\newcommand{\yt}{y_t}

\newcommand{\sigmad}{\sigma_d}
\newcommand{\sigmat}{\sigma_t}

\newcommand{\gamd}{\gamma_d}
\newcommand{\gamt}{\gamma_t}

\newcommand{\rnt}{\rho_t}

\newcommand{\NLO}{\text{NLO}}
\newcommand{\NNLO}{\text{N$^2$LO}}

\newcommand{\sss}{\mathrm{s}}
\newcommand{\ccc}{\mathrm{c}}
\newcommand{\fff}{\mathrm{full}}

\newcommand{\Tgen}{\mathcal{T}}

\newcommand{\TF}{\Tgen_\fff}
\newcommand{\TFf}{\Tgen_{\fff}}

\newcommand{\TFfda}{\TFf^\mathrm{d,a}}
\newcommand{\TFfdb}{\TFf^\mathrm{d,b1}}
\newcommand{\TFfdc}{\TFf^\mathrm{d,b2}}

\newcommand{\KS}{K_\sss}
\newcommand{\KC}{K_\ccc}

\newcommand{\KCd}{\KC^{(d)}}
\newcommand{\KCt}{\KC^{(t)}}

\newcommand{\skrowspace}{0.5em}

\newcommand{\ann}{a_{\text{n--n}}}
\newcommand{\gamnn}{\gamma_{\text{n--n}}}
\newcommand{\rnn}{\rho_{\text{n--n}}}

\begin{document}

\title{Constraints on a possible dineutron state from pionless EFT}

\author{H.-W. Hammer}
\email{Hans-Werner.Hammer@physik.tu-darmstadt.de}
\affiliation{Institut für Kernphysik, Technische Universität Darmstadt, 
64289 Darmstadt, Germany}
\affiliation{ExtreMe Matter Institute EMMI, GSI Helmholtzzentrum 
für Schwerionenforschung GmbH, 64291 Darmstadt, Germany}
\affiliation{Institute for Nuclear Theory, University of Washington,	
Seattle, WA 98195}

\author{Sebastian König}
\email{koenig.389@physics.osu.edu}
\affiliation{Department of Physics, The Ohio State University, Columbus, Ohio 
43210, USA}
\affiliation{Helmholtz-Institut für Strahlen- und Kernphysik (Theorie)\\
and Bethe Center for Theoretical Physics, Universität Bonn, 53115 Bonn,
Germany}

\date{\today}

\begin{abstract}
We investigate the sensitivity of the three-nucleon system to changes in the 
neutron--neutron scattering length to next-to-leading order in the pionless 
effective field theory, focusing on the the triton--$^3$He binding energy 
difference and neutron--deuteron elastic scattering.  Due to the appearance of 
an electromagnetic three-body counterterm at this order, the triton--$^3$He 
binding energy difference remains consistent with the experimental value even 
for large positive neutron--neutron scattering lengths while the elastic 
neutron--deuteron scattering phase shifts are insensitive.  We conclude that 
a bound dineutron cannot be excluded to next-to-leading order in pionless EFT.
\end{abstract}

\maketitle

\section{Introduction}
\label{sec:intro}
The search for dineutron bound states has a long history in physics.  Although 
early experimental searches were negative~\cite{Cohen:1953,Glasgow:1967zz}, 
there has been some evidence for the presence of dineutron configurations in 
the decay of weakly bound nuclei recently.  For example, Bokharev and 
collaborators claim that roughly half of the excited-state decay of $^6$He is 
through the dineutron~\cite{Bochkarev:1985}, Seth and Parker found evidence 
for the presence dineutrons in the breakup of $^5$H, $^6$H, and 
$^8$He~\cite{Seth:1991zz}, and Spyrou~\etal{} observed dineutron emission
in the ground state decay of $^{16}$Be~\cite{Spyrou:2012zz}, to mention a few.
Whether such dineutron configurations correspond to dineutron bound states,
however, is unclear.

In free space the dineutron is believed to be unbound by about $100~\keV$, 
implying a large negative scattering length of about $-20~\fm$.  The most 
precise determination of the neutron-neutron scattering length to date probably 
comes from the final-state interaction in the $\pi^- d$ radiative 
capture reaction~\cite{Gardestig:2005pp}, leading to the value $\ann = -18.63 
\pm 0.27~({\rm expt.}) \pm 0.30~({\rm th.})$~fm~\cite{Chen:2008zzj}.  However, 
the final state interaction peak is expected to be insensitive to the sign of 
$\ann$ such that a positive value of roughly equal magnitude would not be 
excluded~\cite{PhillipsPrivate}.  This issue requires further study. 

Because it is just barely unbound, only a small change in the 
nucleon--nucleon interaction is sufficient to create a bound dineutron.  In 
lattice QCD calculations at unphysically large pion masses of order $800~\MeV$, 
\eg, the spin-singlet nucleon--nucleon system and thus the dineutron is 
bound by about $20~\MeV$~\cite{Yamazaki:2012hi,Beane:2012vq}.  Moreover, a 
relatively small change in the quark masses, as it is discussed in scenarios for 
the variation of fundamental constants, might already be enough to stabilize 
the dineutron.  Kneller and McLaughlin found that big bang nucleosynthesis is 
compatible with dineutron binding energies of up to 2.5 MeV, thus providing 
surprisingly weak constraints~\cite{Kneller:2003ka}.

In the context of the nuclear few-body problem, Witała and Glöckle raised the 
possibility that a slightly bound dineutron might solve some open problems in 
three-body breakup reactions~\cite{Witala:2012te}.  They changed the 
neutron--neutron scattering length by multiplying the CD Bonn potential with an 
overall strength factor ranging from $0.9$ to $1.4$.  One should keep in mind
here that this procedure also changes other low-energy scattering parameters 
besides the scattering length.  Witała and Glöckle found that the 
neutron--deuteron total and differential cross sections do not rule out a bound 
dineutron.  The neutron--neutron final-state interaction configurations 
measured in Ref.~\cite{GonzalezTrotter:2006wz}, however, could not 
simultaneously be reproduced by their rescaled CD Bonn potential.  Overall, a 
dineutron binding energy larger than $100~\keV$ was excluded in their 
study.  These theoretical studies raised interest in new experimental searches 
for the  dineutron.  For example at HIGS/TUNL, there is a proposal to measure 
the neutron--neutron  final state interaction in triton 
photodisintegration~\cite{Tornow:2013,HowellProposal}.

The pionless effective field theory (EFT) is ideally suited to study the 
dependence of low-energy nuclear observables on the neutron--neutron scattering 
length since the latter appears explicitly as a parameter in the theory.  The 
problem of changing other observables as well when rescaling the potential or 
coupling constants is thus avoided.  The theory is applicable for typical 
momenta below the pion mass and is frequently used to describe 
low-energy few-nucleon systems (see e.g. 
Refs.~\cite{Beane:2000fx,Bedaque:2002mn,Epelbaum:2008ga} 
for a reviews and references to earlier work).

Kirscher and Phillips~\cite{Kirscher:2011zn} used pionless EFT to compute a 
model-independent correlation between the difference of the neutron--neutron and 
(Coulomb-modified) proton--proton scattering lengths and the triton--$^3$He 
binding energy difference.  Their calculation was carried out at leading order 
(LO) in the pionless EFT but included isospin breaking effects from the physical
scattering lengths in different charge channels.  They used this correlation to 
differentiate between different measured values of the neutron--neutron 
scattering length and extracted a favored value $\ann=(-22.9\pm 4.1)~\fm$.
They concluded that values outside of this window are not consistent with the 
experimental difference in binding energies between the triton and $^3$He.  
Thus their analysis excludes a bound dineutron.

Here, we carry out a similar analysis focusing on the triton--$^3$He binding 
energy difference to next-to-leading order (NLO).  It was recently shown that a 
new electromagnetic counterterm enters in the pionless EFT at this 
order~\cite{Vanasse:2014kxa,Koenig:2013,Konig:2014ufa}.  Thus the change in the 
binding energy difference between triton and $^3$He can be absorbed by this 
counterterm.  In the next section, we will review some key points of the 
formalism of pionless EFT to NLO and discuss the integral equations for the 
triton and $^3$He systems.  We then present our analysis of three-nucleon 
observables as well as the naturalness of the counterterm and conclude.

\section{Formalism}
\label{sec:formalism}

\paragraph{Effective Lagrangian.}  The effective Lagrangian of pionless EFT can 
be written in the form
\begin{multline}
 \mathcal{L} = N^\dagger\left(\ii D_0+\frac{\vD^2}{2\MN}\right)N
 -d^{i\dagger}\left[\sigmad+\left(\ii D_0+\frac{\vD^2}{4\MN}\right)\right]d^i
 -t^{A\dagger}\left[\sigmat+\left(\ii D_0+\frac{\vD^2}{4\MN}\right)\right]t^A
 \\[0.2cm]
 +\yd\left[d^{i\dagger}\left(N^T P^i_d N\right)+\hc\right]
 +\yt\left[t^{A\dagger}\left(N^T P^A_t N\right)+\hc\right]
 +\mathcal{L}_\mathrm{photon}+\mathcal{L}_3 \,,
\label{eq:L-Nd}
\end{multline}
with the nucleon field $N$ and two dibaryon fields $d^i$ (with spin 1 and
isospin 0) and $t^A$ (with spin 0 and isospin 1), corresponding to the deuteron
and the spin-singlet isospin-triplet virtual bound state in S-wave
nucleon--nucleon scattering.  Spin and isospin degrees of freedom are included
by treating the field $N$ as a doublet in both spaces, but for notational
convenience we usually suppress the spin and isospin indices of $N$.  The
operators $P^i_d$ and $P^A_t$ project out the $^3S_1$ and $^1S_0$
nucleon--nucleon partial waves.  

Furthermore, $\mathcal{L}_\mathrm{photon}$ contains the kinetic and gauge 
fixing terms for the photons, of which we only keep contributions from Coulomb 
photons.  These correspond to a static Coulomb potential between charged 
particles, but for convenience we introduce Feynman rules for a Coulomb-photon 
propagator,
\begin{equation}
 \Delta_{\mathrm{Coulomb}}(k) = \frac{\ii}{\vk^2+\lambda^2} \,,
\label{eq:FR-Coulomb-Propagator}
\end{equation}
which we draw as a wavy line, and factors $(\pm\ii e\,\hat{Q})$ for the
vertices.

In the spin-doublet S-wave channel where the triton and $^3$He reside,
a three-body contact interaction is required for renormalization already at 
leading order in the EFT~\cite{Bedaque:1999ve}.  We write it here in the form 
given by Ando and Birse~\cite{Ando:2010wq},
\begin{equation}
 \mathcal{L}_3=\frac{\MN H(\Lambda)}{3\Lambda^2}N^\dagger\bigg(\yd^2\,
 (d^i)^\dagger d^j \sigma^i \sigma^j+\yt^2\,(t^A)^\dagger t^B \tau^A\tau^B
 -\yd\yt[(d^i)^\dagger t^A \sigma^i \tau^A +\hc]
 \bigg)N \,,
\label{eq:L-3}
\end{equation}
where $\sigma^i$ and $\tau^A$ are Pauli matrices in spin and isospin
space, $\Lambda$ is a momentum cutoff applied in the three-body equations
discussed below and $H(\Lambda)$ a known log-periodic function of the cutoff
that depends on a three-body parameter $\Lambda_*$~\cite{Bedaque:1999ve}.

\begin{figure}[htbp]
\centering
\includegraphics[clip,width=\textwidth]{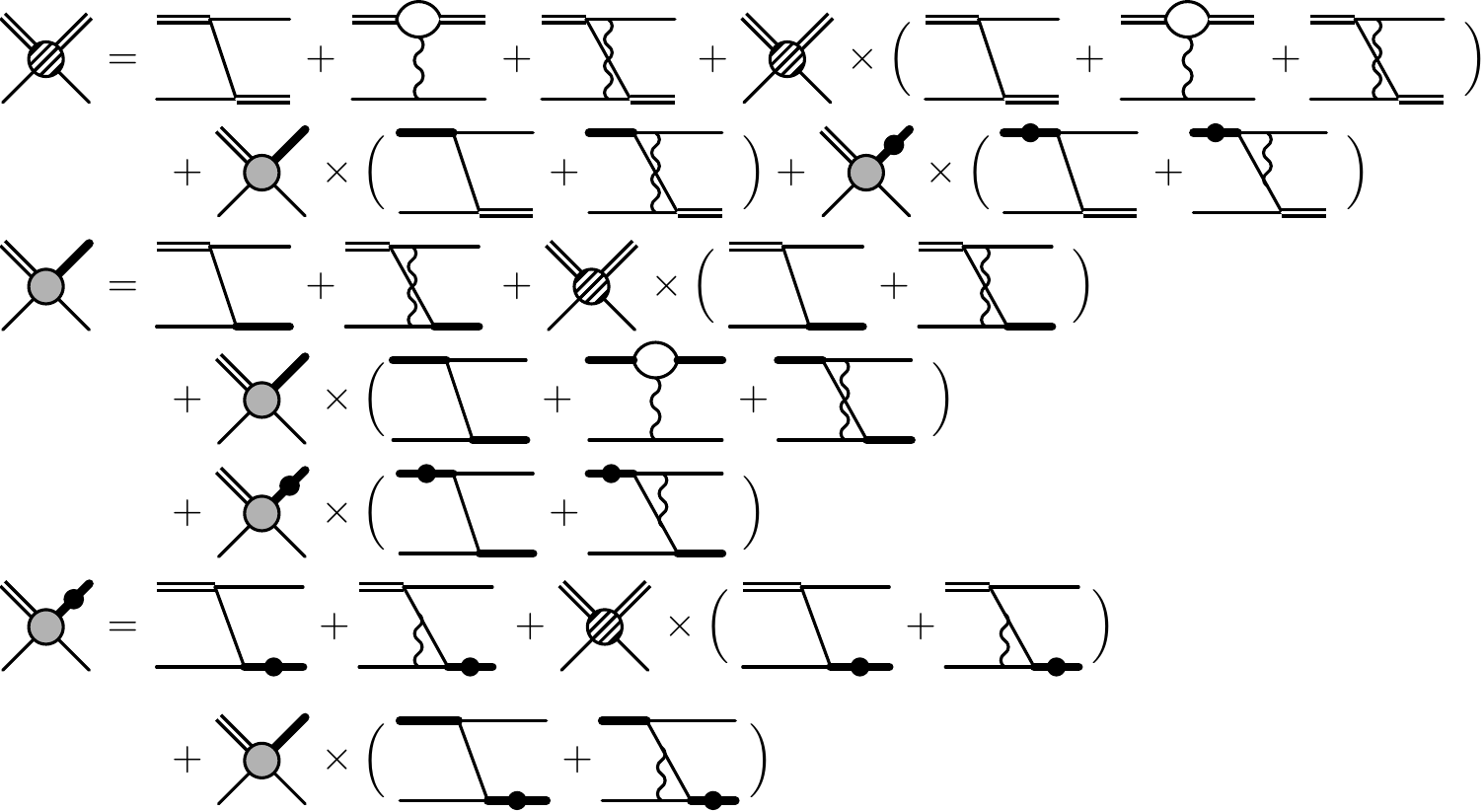}
\caption{Coupled-channel integral equation for the full scattering amplitude 
in the $^3$He channel.  Thin single lines represent nucleons, the double line 
stands for a deuteron, and the thick lines for dibaryons in the spin-singlet 
state (with an additional dot to indicate the $p$--$p$ channel
which is treated separately).  The diagrams representing the three-nucleon 
force have been omitted.}
\label{fig:pd-IntEq-full}
\end{figure}

\paragraph{Scattering equation.}  The integral equation for the 
proton-deuteron scattering amplitude in the $^3$He channel is displayed 
diagrammatically in Fig.~\ref{fig:pd-IntEq-full}.  It is a three-component 
quantity that we denote as $\TF=(\TFfda,\TFfdb,\TFfdc)^T$, where all three 
components are in general functions of the total energy $E$ as well as of the 
in- and outgoing momenta $k$ and $p$, \ie, $\Tgen=\Tgen(E; k, p)$.  Everything 
is projected onto the S-waves here such that there is no angular dependence.

Using the formal notation 
\begin{equation}
 A \otimes B \equiv \frac1{2\pi^2}
 \int_0^\Lambda\dd q\,q^2\,A(\ldots,q)B(q,\ldots)
\end{equation}
and the abbreviations $g_{dd,tt}=\MN y{d,t}^2/2$, $g_{dt}=\MN\yd\yt/2$, the 
equation can be written as
\begin{multline}
\allowdisplaybreaks
 \begin{pmatrix}\TFfda\\[\skrowspace]\TFfdb\\[\skrowspace]\TFfdc\end{pmatrix}
 = \begin{pmatrix}
 g_{dd}\left(\KS+\frac{2H(\Lambda)}{\Lambda^2}\right)\\[\skrowspace]
 -g_{dt}\left(\KS+\frac{2H(\Lambda)}{3\Lambda^2}\right)\\[\skrowspace]
 -g_{dt}\left(2\KS+\frac{4H(\Lambda)}{3\Lambda^2}\right)
 \end{pmatrix}
 + \begin{pmatrix}g_{dd}\left(\KCd+K_{\text{box}}\right)\\[\skrowspace]
 -g_{dt}K_{\text{box}}\\[\skrowspace]
 -2g_{dt}K_{\text{tri}}^{(\text{in})}\end{pmatrix} \\
 +\begin{pmatrix}
 -g_{dd}D_d\left(\KS+\frac{2H(\Lambda)}{\Lambda^2}\right) &
 g_{dt}D_t\left(3\KS+\frac{2H(\Lambda)}{\Lambda^2}\right) & 0\\[\skrowspace]
 g_{dt}D_d\left(\KS+\frac{2H(\Lambda)}{3\Lambda^2}\right) & 
 g_{tt}D_t\left(\KS-\frac{2H(\Lambda)}{3\Lambda^2}\right) & 0\\[\skrowspace]
 g_{dt}D_d\left(2\KS+\frac{4H(\Lambda)}{3\Lambda^2}\right) &
 -g_{tt}D_t\left(2\KS+\frac{4H(\Lambda)}{3\Lambda^2}\right) & 0
 \end{pmatrix}
 \otimes \begin{pmatrix}
 \TFfda\\[\skrowspace]\TFfdb\\[\skrowspace]\TFfdc\end{pmatrix} \\
 + \begin{pmatrix}
 -g_{dd}D_d\left(\KCd+K_{\text{box}}\right) &
 3g_{dt} D_t K_{\text{box}} & g_{dt}D_t^{pp}
 \left(3\KS + 3K_{\text{tri}}^{(\text{out})}
 +\frac{2H(\Lambda)}{\Lambda^2}\right)\\[\skrowspace]
 g_{dt} D_d K_{\text{box}} &
 -g_{tt}D_t\left(\KCt-K_{\text{box}}\right) & -g_{tt}D_t^{pp}
 \left(\KS + K_{\text{tri}}^{(\text{out})}
 +\frac{2H(\Lambda)}{3\Lambda^2}\right)\\[\skrowspace]
 2g_{dt} D_d K_{\text{tri}}^{(\text{in})} &
 -2g_{dt} D_t K_{\text{tri}}^{(\text{in})} &
 -g_{tt}D_t^{pp}\times\frac{4H(\Lambda)}{3\Lambda^2}
 \end{pmatrix} \\
 \otimes\begin{pmatrix}
 \TFfda\\[\skrowspace]\TFfdb\\[\skrowspace]\TFfdc
 \end{pmatrix} \,,
\label{eq:pd-IntEq-D-full-more}
\end{multline}
where we have now omitted all arguments for brevity.  At leading order,
\begin{equation}
 D_{d,t}(E;q)
 = -\frac{4\pi}{\MN y_{d,t}^2} \times \frac{1}
 {-\gamma_{d,t}+\sqrt{3q^2/4-\MN E-\ii\eps}}
\label{eq:D-LO}
\end{equation}
with $\gamd\equiv\sqrt{\MN E_d}$ and $\gamt\equiv1/a_t$.  In the $p$--$p$ 
channel, one has the modified
propagator~\cite{Ando:2010wq,Kong:1998sx,Kong:1999sf}
\begin{multline}
 D_{t,pp}(E;q)
 = -\frac{4\pi}{\MN y_{t}^2} \times \frac{1}
 {-1/a^C-\alpha\MN\left(\psi(\ii\eta)+\frac{1}{2\ii\eta}-\log(\ii\eta)\right)}
 \\ \mathtext{with} 
 \eta = \alpha\MN/2\times\left(3q^2/4-\MN E-\ii\eps\right)^{-1/2} \,.
\label{eq:D-LO-pp}
\end{multline}
Explicit expressions for the kernel functions---$K_s(E;k,p)$, $K_c^{(d,t)}(E; 
k, p)$, etc.---as well as a more detailed derivation of 
Eq.~\eqref{eq:pd-IntEq-D-full-more} can be found in Ref.~\cite{Konig:2014ufa}, 
Secs.~III and in particular V.B.2.

\paragraph{Higher-order corrections.}  At next-to-leading order there are 
perturbative corrections to the propagators in Eqs.~\eqref{eq:D-LO} 
and~\eqref{eq:D-LO-pp} which are linear in the corresponding effective ranges 
(\cf~Ref.~\cite{Vanasse:2014kxa} and, for an expression in the same notation 
used here, in particular Eq.~(3) in Ref.~\cite{Konig:2013cia}).  Our fully 
perturbative \NLO{} calculation is based on Refs.~\cite{Vanasse:2013sda} 
and~\cite{Vanasse:2014kxa}.

\paragraph{The triton channel.}  We only show the integral equation for the 
$^3$He ($p$--$d$ doublet) channel explicitly.  It is straightforward to obtain 
the integral equation for the $^3$H ($n$--$d$ doublet) channel from 
Eq.~\eqref{eq:pd-IntEq-D-full-more}.  As a first step to that end one simply 
removes all kernel functions in Eq.~\eqref{eq:pd-IntEq-D-full-more} that 
correspond to Coulomb-photon exchanges ($K_c^{d,t}$, $K_{\text{box}}$, 
$K_{\text{tri}}^{(\text{in/out})}$).  If one furthermore lets 
$D_t^{pp}\rightarrow D_t$, one obtains just the scattering equation for the 
$n$--$d$ doublet amplitude in the isospin limit, which can actually be reduced 
to a two-channel equation~\cite{Konig:2014ufa}.  In this work, however, we want 
to study the effect of varying the neutron--neutron scattering length, so we 
rather let $D_t^{pp}\rightarrow D_t^{nn}$, where
\begin{equation}
 D_{t,nn}(E;q)
 = -\frac{4\pi}{\MN y_{t}^2} \times \frac{1}
  {-\gamnn+\sqrt{3q^2/4-\MN E-\ii\eps}} \,.
\label{eq:D-LO-nn}
\end{equation}
For $\ann<0$, we simply set $\gamnn\equiv1/\ann$.  In the regime of positive 
$\ann$, corresponding to the existence of a hypothetical bound dineutron, it is 
more convenient to match the propagator to the effective range expansion around 
the dineutron pole.  Accordingly, we set
\begin{equation}
 \gamnn \equiv \frac{1}{\ann}\left(1+\frac{\rnn}{2\ann}\right)
 \mathtext{for} \ann>0 \,,
\end{equation}
where for simplicity we assume $\rnn=\rnt=2.73~\fm$~\cite{Beane:2000fx}.  At 
leading order, this only corresponds to a constant offset and does not 
otherwise affect the result.  The effect of varying $\rnn$ away from the 
isospin-symmetric case at \NLO{} will be discussed below.

\section{Results}
\label{sec:results}

From Eq.~\eqref{eq:pd-IntEq-D-full-more} and its analog for the 
neutron--deuteron case one can extract both scattering information---for 
example the $n$--$d$ doublet scattering length which we use as physical input 
to fix the three-nucleon force $H(\Lambda)$---and bound state properties.  To 
extract the binding energies of the triton and $^3\mathrm{He}$, we look 
for poles (as a function of the energy) in the corresponding scattering 
amplitudes at negative energies.  In practice, this can simply be done by 
studying the homogeneous versions of Eq.~\eqref{eq:pd-IntEq-D-full-more} and 
its analog for the triton channel.

\subsection{Leading order}

\begin{figure}[htbp]
\centerline{
\includegraphics[clip, width=0.48\textwidth]{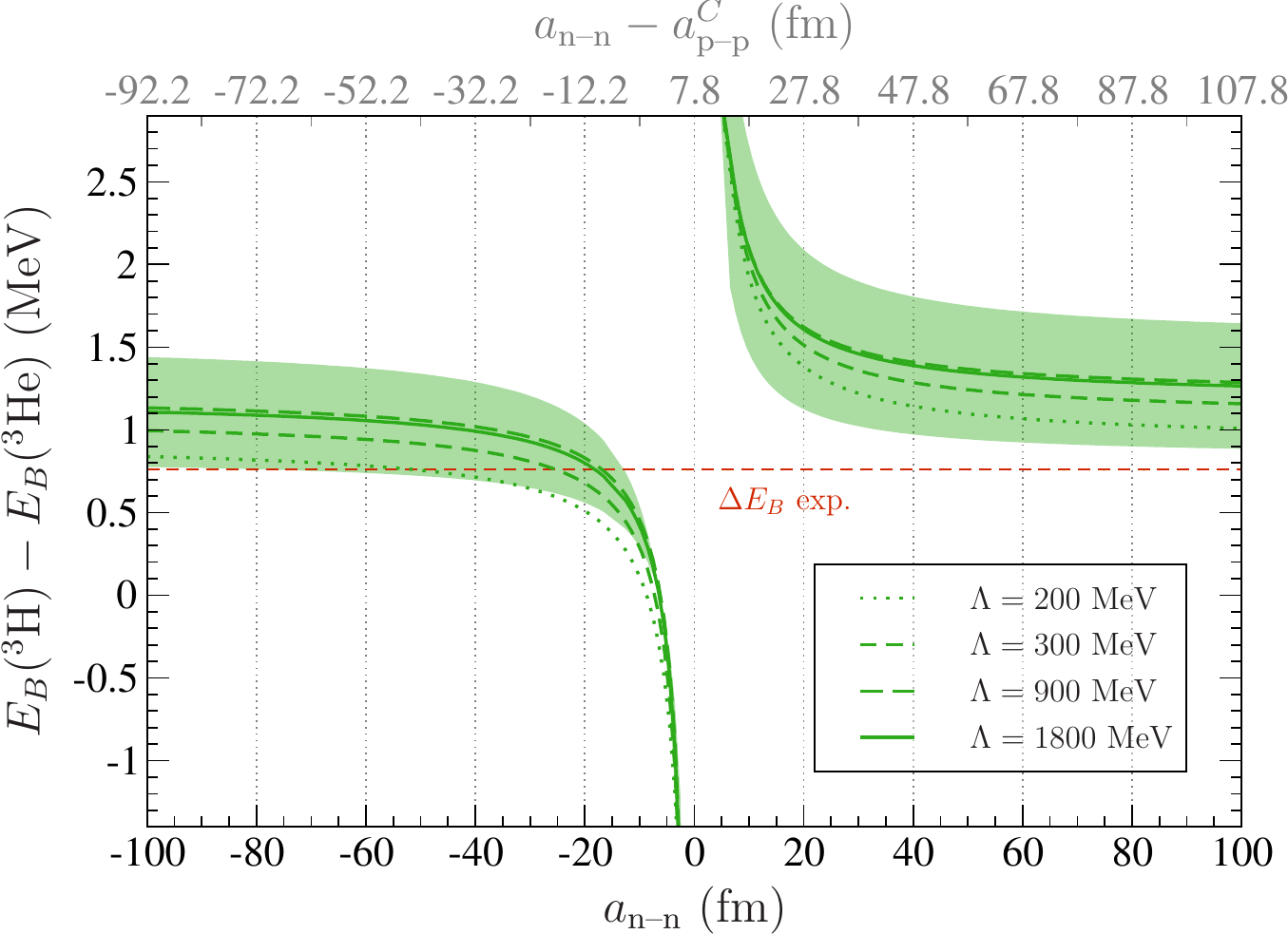}
\quad
\includegraphics[clip, width=0.48\textwidth]{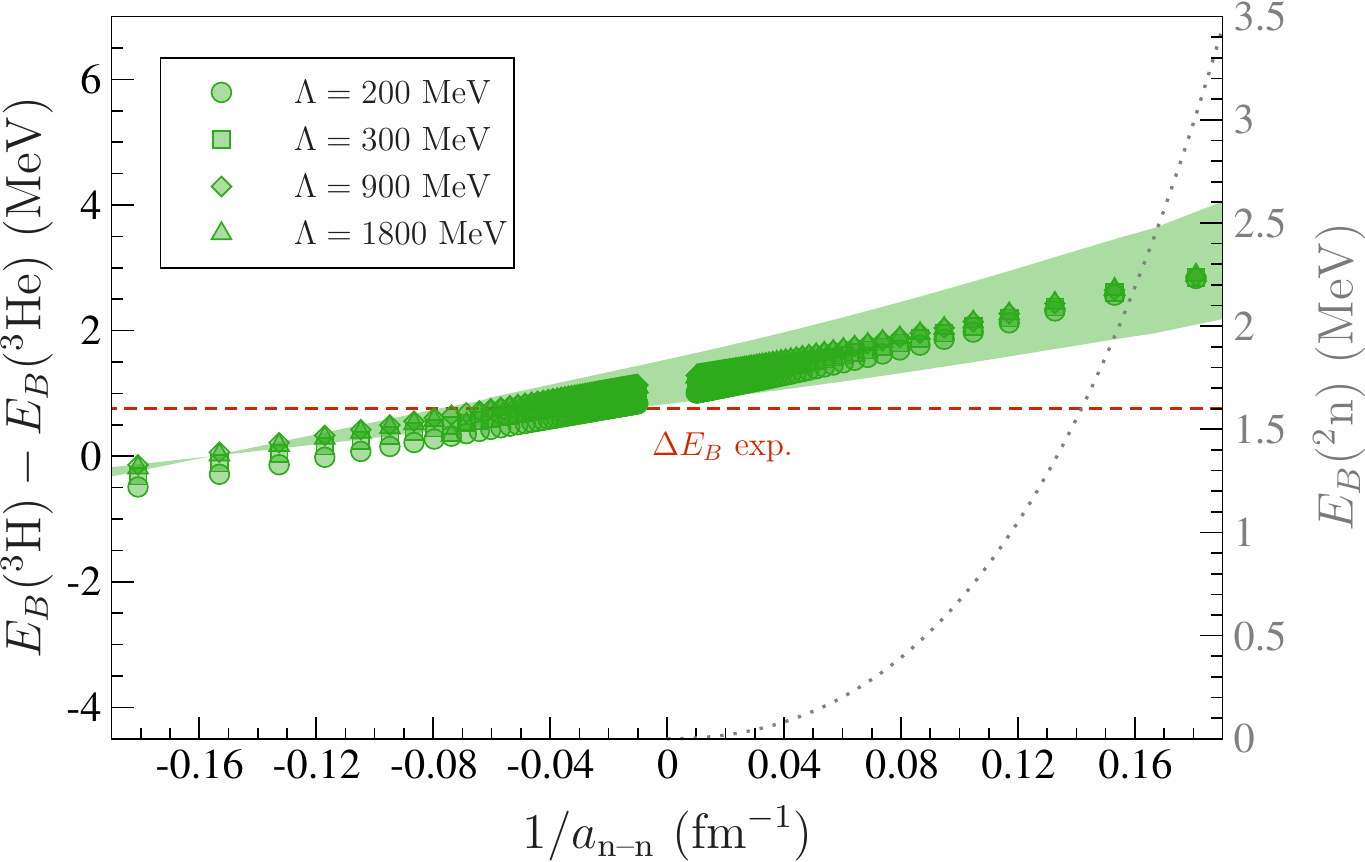}
}
\caption{ Leading-order triton--$^3\mathrm{He}$ binding energy difference as a
function of the neutron--neutron scattering length (left panel) and as a 
function of the inverse neutron--neutron scattering length (right panel).  The 
shaded bands were generated by varying the $\Lambda=1800~\MeV$ within $\pm30\%$. 
The dotted line in the right panel shows the (hypothetical) dineutron energy as 
a function of $\ann^{-1}$.}
\label{fig:DeltaE-a-LO}
\end{figure}

In the left panel of Fig.~\ref{fig:DeltaE-a-LO}, we show the 
triton--$^3\mathrm{He}$ binding energy difference
\begin{equation}
 \Delta E_3 = E_B(^3\mathrm{H}) - E_B(^3\mathrm{He})
\end{equation}
as a function of the $n$--$n$ scattering length $\ann$, both for negative and 
for positive values of the latter quantity.\footnote{We note that the 
region of small scattering lengths, $\ann\approx 0~\fm$, should be discarded.
This region is clearly excluded by experiment.  Moreover, our theory requires a 
scattering length large compared to the range of the interaction.}

In the regime of negative $\ann$ our results agree nicely with the findings of 
Kirscher and Phillips~\cite{Kirscher:2011zn}, who calculated the binding-energy 
difference using the resonating group method (RGM).  Those authors did not 
explore the possibility of a (large) positive $n$--$n$ scattering length.  
Consequently, only the negative arm of the pole at $\ann=0~\fm$ was found in 
Ref.~\cite{Kirscher:2011zn}, and from that one might naïvely think that positive 
values of $\ann$ are clearly excluded.  However, the relevant quantity here is 
not actually $\ann$, but rather its inverse.  In the right panel of 
Fig.~\ref{fig:DeltaE-a-LO}, we show that $\Delta E_3$ is indeed a continuous 
function of $\ann^{-1}$ around $\ann^{-1}\approx 0$.

This raises the question of how well one can determine $\ann$ from a 
leading-order pionless EFT calculation.  In the left panel of 
Fig.~\ref{fig:DeltaE-a-LO}, we show an error band that was generated by 
varying the (essentially cutoff-converged) $\Lambda=1800~\MeV$ curve within 
$\pm30\%$, corresponding roughly to the estimated size of an \NLO~correction in 
pionless EFT.  From that, a positive value of $\ann$ is just barely excluded, 
and by making a slightly more conservative estimate one would find that such a 
case can be marginally consistent with the physical binding-energy difference.

One could argue now that our band might be overestimating the uncertainty since 
some contributions can be expected to cancel in the difference of 
the---individually calculated---$^3\mathrm{H}$ and $^3\mathrm{He}$ energies.  
However, it has recently been shown~\cite{Vanasse:2014kxa} that a new three-body
counterterm $H_{0,1}^{(\alpha)}(\Lambda)$ is necessary to renormalize the 
doublet-channel $p$--$d$ system at next-to-leading order.  This might mean that 
the actual uncertainty is somewhat larger than what one might expect based on 
the considerations in Ref.~\cite{Kirscher:2011zn}.  We will investigate this
issue in the next subsection.

\subsection{Next-to-leading order}

\begin{figure}[htbp]
\centerline{
\includegraphics[clip,width=0.48\textwidth]{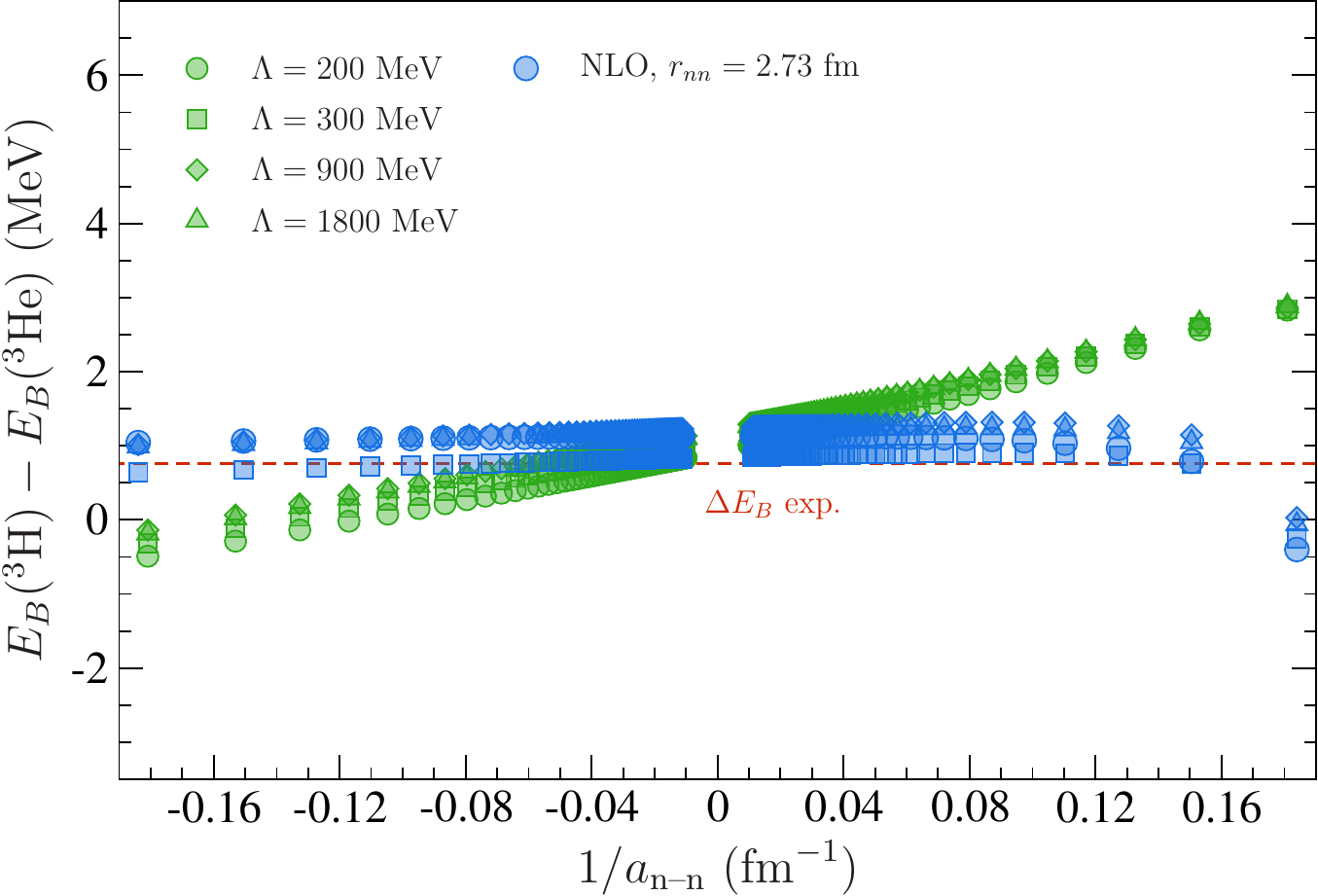}
\quad
\includegraphics[clip,width=0.48\textwidth]{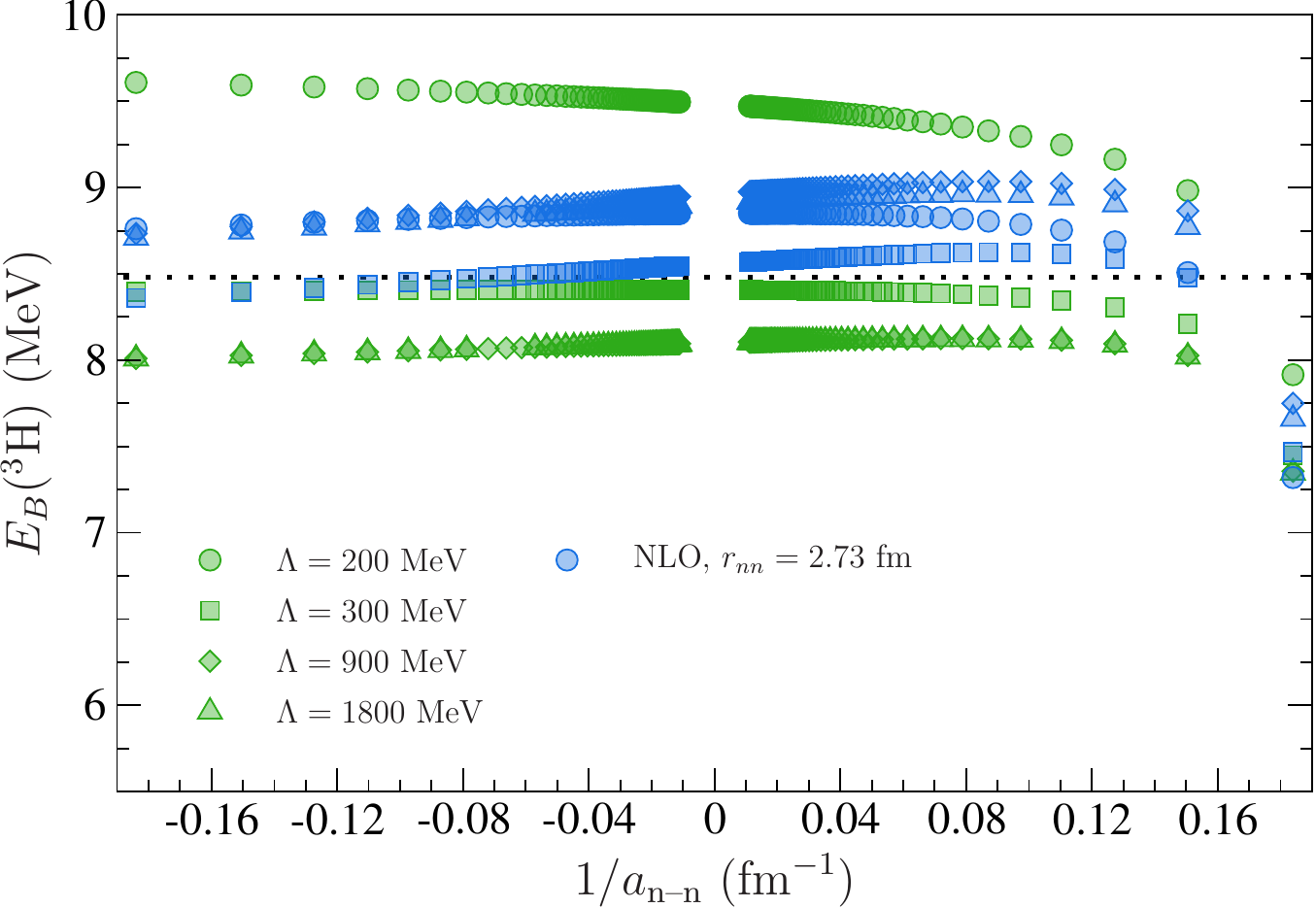}
}
\caption{Left panel: \NLO~result for the triton--$^3\mathrm{He}$ binding 
energy difference as a function of the inverse neutron--neutron scattering 
length. Right panel: \NLO~result for the triton binding energy as a 
function of the inverse neutron--neutron scattering length.}
\label{fig:DeltaE-aInv-NLO}
\end{figure}

When the new counterterm is fit to reproduce the experimental $^3\mathrm{He}$ 
binding energy, one can no longer predict both $E_B(^3\mathrm{H})$
and $E_B(^3\mathrm{He})$.  Still, there is a dependence of $\Delta E_3$ on 
$\ann$ that comes from the triton binding energy.  The result of this 
calculation is shown in the left panel of Fig.~\ref{fig:DeltaE-aInv-NLO}.

The shape of the curves is now different compared to the leading-order result, 
and it looks now as if not only a large range of negative $\ann$ would be 
consistent with the experimental $\Delta E_3$, but also some positive values 
would in fact be allowed.  If one looks directly at the prediction for the 
triton binding energy, as shown in the right panel of
Fig.~\ref{fig:DeltaE-aInv-NLO}, one finds that the results at leading order and 
next-to-leading order nicely overlap and that the cutoff-dependence is smaller 
at \NLO.  To make the figure less cluttered, we show no explicit error bands 
here.  Varying the $n$--$n$ effective range by $\pm10\%$ around
$\rnn=\rnt=2.73~\fm$ moves the NLO curves in Fig.~\ref{fig:DeltaE-aInv-NLO}
up and down by about $0.1$-$0.15~\MeV$.  The influence of this parameter 
therefore quite small, but it makes it a little bit harder yet to exclude a 
bound dineutron on the grounds of pionless effective field theory.

With the new three-body counterterm present one can no longer make a 
parameter-free prediction for $\Delta E_3$ and next-to-leading order.  
Unfortunately, fitting $H_{0,1}^{(\alpha)}(\Lambda)$ to the doublet-channel 
$p$--$d$ scattering length does not restore much predictive power since that 
quantity is surprisingly poorly known~\cite{Black:1999ab}.

\begin{figure}[htbp]
\centering
\includegraphics[clip, width=0.6\textwidth]{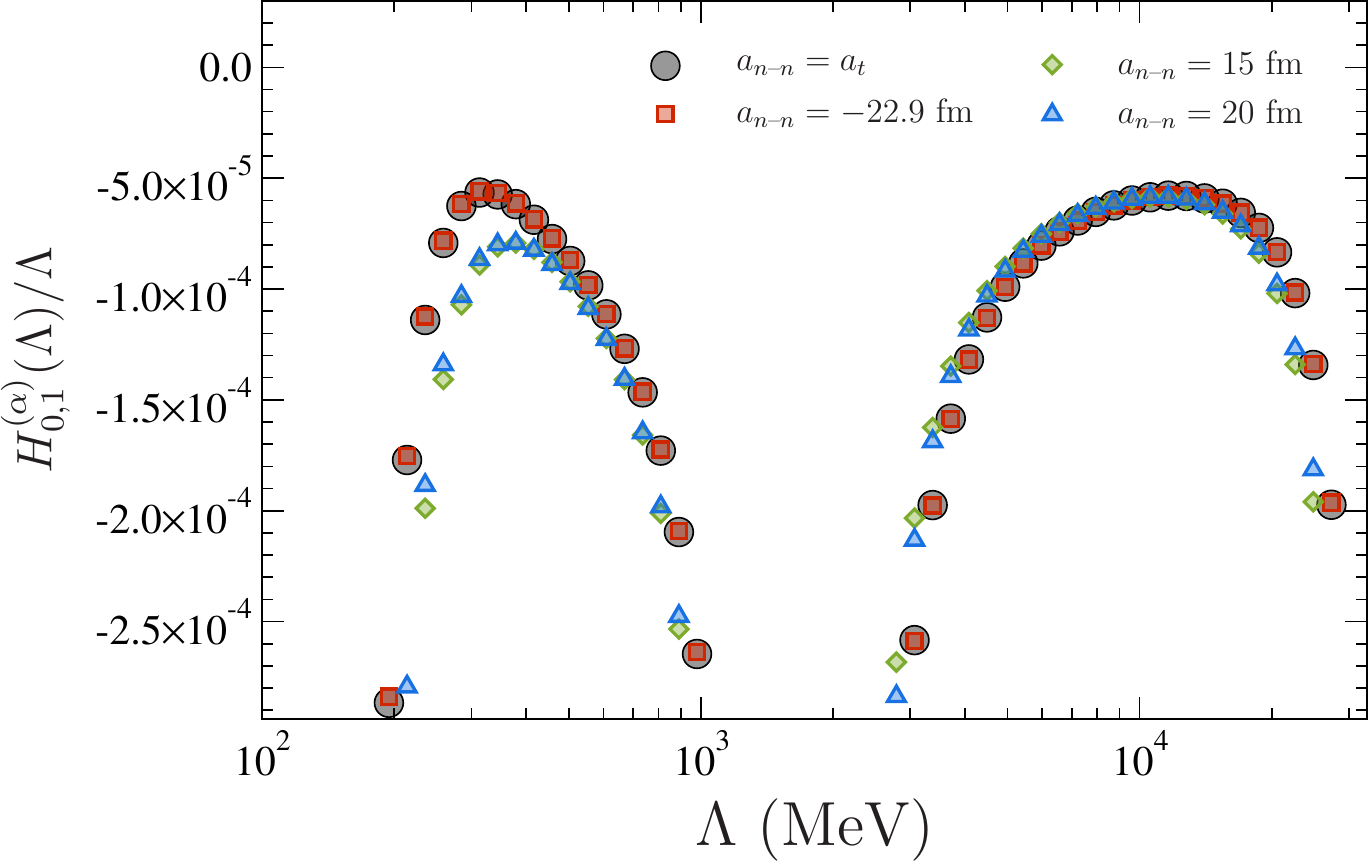}
\caption{Electromagnetic counterterm $H_0^{(\alpha)}(\Lambda)$ as a function
of the cutoff $\Lambda$ for different neutron-neutron scattering lengths.}
\label{fig:Ha1-ann}
\end{figure}

This makes it interesting to see what can be learned from the $\ann$-dependence 
of the new counterterm itself.  In particular, one can check if it becomes 
unnatural for positive values of $\ann$.  In Fig.~\ref{fig:Ha1-ann} we answer 
this question in the negative.  The leading behavior is~\cite{Vanasse:2014kxa}
\begin{equation}
 H_{0,1}^{(\alpha)}(\Lambda) \propto \Lambda \,,
\end{equation}
with subleading logarithmic corrections.  One clearly sees that the coefficient 
does not change its order of magnitude if one considers positive values of 
$\ann$, neither for natural---$\OO(m_\pi)$---nor for asymptotically large 
cutoffs.  On this ground we conclude that one cannot rule out the existence of a 
shallow bound $n$--$n$ state from pionless EFT at NLO.  It might thus be 
worthwhile to continue investigating that possibility, both theoretically and 
experimentally.

\begin{figure}[htbp]
\centering
\includegraphics[clip, width=0.6\textwidth]{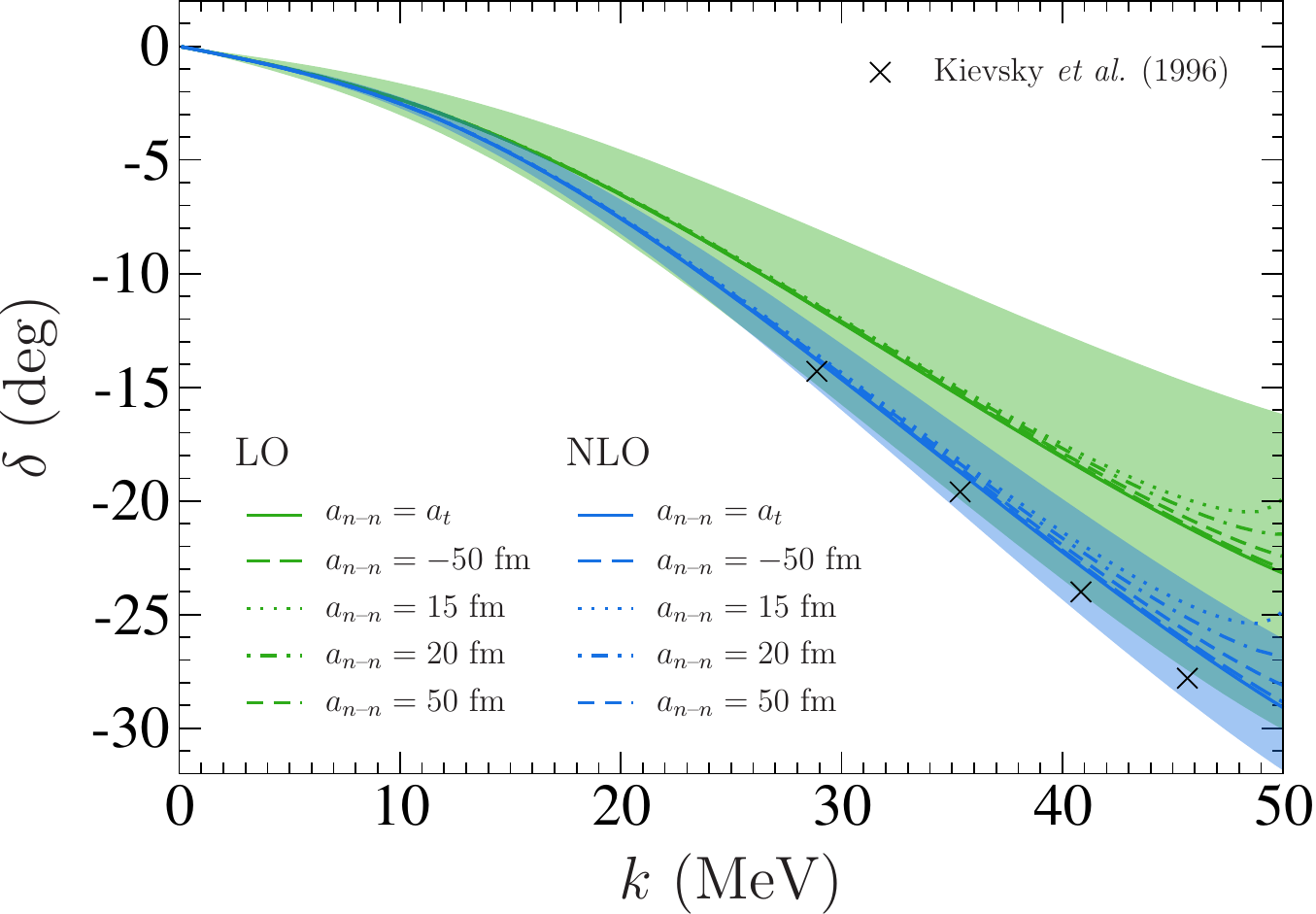}
\caption{S-wave $n$--$d$ doublet channel scattering phase shifts as functions 
of the center-of-mass momentum $k$ for a cutoff $\Lambda=1800~\MeV$ and several 
values of the $n$--$n$ scattering length (see plot legend).  The shaded bands 
were generated by letting the $\ann=a_t$ curve vary within $\pm30\%$ (LO) and 
$\pm10\%$ (NLO).  The crosses are the results from the AV18+UR potential-model 
calculation reported in Ref.~\cite{Kievsky:1996ca}.}
\label{fig:Phase-D-nn-NLO}
\end{figure}

Finally, we show in Fig.~\ref{fig:Phase-D-nn-NLO} that the $n$--$d$ scattering 
phase shifts are quite insensitive to variations of $\ann$ to (large) positive 
values.  We have only included S-wave phase shifts here since the P-wave result 
looks very similar and there is even less effect in higher partial waves.  Thus 
there is very little sensitivity in elastic scattering observables to changes 
from $\ann^{-1}$ from small negative to small positive values which is
in agreement with the findings of Witała and Glöckle~\cite{Witala:2012te}.

\section{Conclusion}

We conclude that a bound dineutron cannot be excluded based on the 
triton--$^3$He binding energy difference and elastic scattering results in 
pionless EFT at \NLO.  This result provides support for planned dineutron 
searches by measuring the neutron--neutron final-state interaction in triton
photodisintegration~\cite{Tornow:2013,HowellProposal}.  Even if no bound 
dineutron is found such experiments will be useful to settle the controversy 
about the value of the neutron--neutron scattering length 
(\cf~Ref.~\cite{Kirscher:2011zn}).  We note in passing that the existence of a 
bound dineutron would also provide a way to understand the recent data by the 
HypHI collaboration suggesting a bound $nn\Lambda$ 
system~\cite{Rappold:2013jta}.  Hiyama and collaborators showed that such a 
bound state could be accommodated if a bound dineutron state
existed~\cite{Hiyama:2014cua}.  The resulting shift in the 
triton--$^3\mathrm{He}$ binding energy difference can be absorbed by a 
naturally-sized NLO three-body force $H_{0,1}^{(\alpha)}$ as demonstrated 
above.  We note, however, that a bound dineutron would be difficult to 
accommodate in standard approaches to charge-symmetry breaking in the 
two-nucleon system~\cite{Gardestig:2009ya,Miller:1994zh}.  This issue requires 
further theoretical study.

It would also be valuable to extend the calculation to \NNLO{} and to 
investigate whether the neutron--neutron final-state interaction configurations 
measured in Ref.~\cite{GonzalezTrotter:2006wz} can simultaneously be reproduced 
with a positive scattering length.  The latter would require a calculation of 
deuteron breakup reactions and is beyond the scope of this work.

\begin{acknowledgments}
We thank D.~R.~Phillips for comments on the manuscript and W.~Tornow for 
providing Ref.~\cite{HowellProposal}.  We also thank the Institute for Nuclear 
Theory at the University of Washington for its hospitality and the Department of 
Energy for partial support during the completion of this work.  Furthermore, 
this research was supported in part by the NSF under Grant No. PHY--1306250, by 
the NUCLEI SciDAC Collaboration under DOE Grant DE-SC0008533, by the DFG 
(SFB/TR 16 ``Subnuclear Structure of Matter''), by the BMBF under grant 
05P12PDFTE, and by the Helmholtz Association under contract HA216/EMMI.
\end{acknowledgments}

\end{document}